\documentclass[amssymb,amsmath,prl,twocolumn,preprintnumbers,superscriptaddress,nofootinbib]{revtex4-2}
\usepackage{graphicx}
\usepackage{bm}
\usepackage{amssymb,amsmath}
\usepackage{mathrsfs}
\usepackage{latexsym}
\usepackage{color}
\usepackage[normalem]{ulem} 
\usepackage{dcolumn}
\usepackage[colorlinks=true,citecolor=blue,urlcolor=blue]{hyperref}
\usepackage[usenames,dvipsnames]{xcolor}
\usepackage{xspace}
\usepackage{graphicx}

\usepackage{multirow}
\usepackage{xfp} %%% for basic arithmetic within tex
\usepackage{comment}
\usepackage{nameref}
\usepackage{lineno}
\newcommand{\comicpage}[1]{}

\newcommand{\com}[1]{{\sf\color[rgb]{0,0,1}{[#1]}}}
\newcommand{\old}[1]{{\color[rgb]{0.5,0.5,0.5}\sout{#1}}}

\newcommand{\rhoc}{\rho_{\rm c}}

\newcommand{\NS}{\mathrm{NS}}
\newcommand{\BH}{\mathrm{BH}}
\newcommand{\Hz}{\mathrm{Hz}}
\newcommand{\tide}{\mathrm{tide}}
\newcommand{\orb}{\mathrm{orb}}
\newcommand{\GW}{\mathrm{GW}}
\newcommand{\e}[1]{\times 10^{#1}}

\usepackage[english]{babel}
\usepackage{xspace}
\usepackage{color,soul}

\newcommand{\quant}[1]{\left(#1\right)}
\newcommand{\quantb}[1]{\left[#1\right]}
\newcommand{\abs}[1]{\left| #1 \right|}
\providecommand{\aj}{AJ}
\providecommand{\apj}{ApJ}

\providecommand{\mnras}{MNRAS}
\providecommand{\nat}{Nature}

\newcommand{\UIUC}{\affiliation{Illinois Center for Advanced Studies of the Universe, Department of Physics, University of Illinois at Urbana-Champaign, \\Urbana, Illinois, 61801, USA}}

\renewcommand{\com}[1]{}

\renewcommand{\old}[1]{}
\newcommand{\mainEqTorque}{Eq.~\eqref{eq:tidal-torque-unreduced}}

\begin{document}
%% Body of the Letter: title block, abstract and all main text through the
%% End Matter.  No preamble, no \begin{document}, no \bibliography -- those
%% belong to the wrappers, so this file is shared verbatim by:
%%   main.tex              (Letter + Supplemental Material, for arXiv)
%%   letter-standalone.tex (Letter alone, for the journal submission)

\title{
Neutron Star--Primordial Black Hole Binaries: \\ Dynamical Tides, Resonance Locking, and Gravitational Waves
}

\author{Isaac Legred}\email{ilegred2@illinois.edu} \UIUC
\author{Nicol\'as Yunes} \UIUC

\begin{abstract}

The orbital evolution of a neutron star and a much lower-mass primordial black hole can differ strikingly from that of typical compact binaries. 
Dynamical tides, analogous to those in the Solar System's giant planet-moon systems, can accelerate inspiral or induce outspiral, depending on the orbital configuration. 
We show how gravitational-wave emission mediates this evolution when neutron star modes remain locked in resonance with the orbit. 
These effects may leave observable signatures in pulsar timing or gravitational waves.

\end{abstract}

\maketitle

\noindent {\bf{\textit{Introduction.}}} Compact binary inspirals are generally studied using the well-understood model of two structureless point masses in general relativity~\cite{Peters:1963ux}.  Effects that go beyond this point-particle model, such as matter effects  in neutron stars, are treated as small perturbations~\cite{Hartle:1968si, Flanagan:2007ix, Hinderer:2007mb, Yagi:2013awa,Abac:2023ujg,Yu:2025ptm, Gamba:2023mww}.  The motivation for this perturbative treatment is the exceptionally strong scaling of equilibrium tidal effects with orbital separation ($r^{-5}$). Beyond a certain orbital separation, tidal effects are assumed to be completely negligible in compact binary inspirals.  However, the weakness of tidal effects is not a property of relativity, but rather simply Newtonian gravity~\cite{Zahn:1977mi, Press:1977} applied to compact stars. 

The difficulty of explaining the rapid orbital migration of Jupiter and Saturn’s moons with equilibrium tides~
\cite{Lainey:2009, Lainey:2012} motivated~\citet{Fuller:2016} to consider \textit{dynamical tides}. Dynamical tides arise when a companion’s time-dependent tidal field excites internal waves or oscillation modes in an extended body. When the tidal forcing frequency approaches a mode frequency, resonant coupling can enhance the exchange of energy and angular momentum between the body and the orbit, producing a large response even if the tidal field is weak. In giant planet–moon systems, this enhanced coupling can transfer angular momentum from the planet’s rotation to the moon’s orbit and drive rapid outward migration.

The obvious limitation of explaining secular orbital changes using dynamical tides is that the large effects of resonant energy transfer are only active so long as the mode remains resonantly coupled to the orbit.  If the orbit changes, and the internal mode frequency does not, then the resonance is lost.  However, in certain cases, stable equilibrium configurations in the phase space of orbital parameters and internal dynamical degrees of freedom appear, which prevent large departures of the orbital frequency from the mode frequency.  This process is termed \emph{resonance locking}~\cite{Witte:1999tq}.  

Once resonance locking occurs, the mode and the orbit evolve together as the mode frequency changes. In the case of Jupiter's and Saturn's moons, this evolution is driven by changes in the giants' internal structure. In the case of comparable-mass neutron-star binaries, earlier calculations that neglected nonlinear hydrodynamics found that gravitational-wave inspiral is too rapid for tidal spin evolution to maintain resonance~\cite{Ho:1998hq,Burkart:2013jua}. More recently, however, calculations including nonlinear hydrodynamics have found that the resulting amplitude-dependent shifts of low-order neutron-star g-mode frequencies may allow a mode to remain near resonance with the tidal forcing as the inspiral proceeds~\cite{Kwon:2024zyg,Kwon:2025zbc}.  Nonetheless, it remains unclear to what extent double neutron star binaries can have their evolution substantially modified by resonance-locked energy transfer.

How does resonance locking then affect compact binary inspiral? Resonance locking in inspiraling double white dwarf binaries was studied by Refs.~\cite{Burkart:2013, Burkart:2013jua}. Because the stellar moments of inertia are small compared with the orbital moment of inertia, the orbital dynamics are only weakly modified by resonance locking, while the spins evolve on the gravitational-wave timescale. While neutron stars have very different internal physics than giant planets (or even white dwarfs), their coupled spin and orbital evolution during an established resonance lock is constrained by angular-momentum (and energy) balance. The internal physics  determines only whether that lock can be established and maintained.

For resonance locking to occur, tidal oscillation modes must have time to build sufficient amplitude near resonance. In relativistic binaries, this requires the gravitational-wave driven inspiral to be sufficiently slow. For comparable-mass binaries (such as double white dwarfs), the inspiral is slower at large orbital separations, where the orbital dynamics are approximately Newtonian. For extreme mass-ratio binaries, inspiral is also slow even at smaller separations because gravitational waves remove only a small fraction of the orbital angular momentum per dynamical time.

In this Letter, we investigate the orbital dynamics of a neutron star in a binary with a much lower mass compact companion. We find that such a binary can lead to diverse and strikingly different binary behavior than for comparable-mass binaries on the gravitational-wave inspiral timescale.  We first consider the necessary conditions in a neutron-star/compact companion binary to establish resonance locking.  We then move onto the dynamics of a system that maintains resonance locking, and compute the modifications to the first post-Newtonian equations of motion. We continue by considering the long-term observable implications of resonance locking, and identify a special orbital binary configuration where resonance locking must break down. We finally comment on the astrophysical implications of resonance locking becoming active for the possible case of low-mass primordial black hole that inspirals into a neutron star.  If primordial black holes with masses smaller than $\sim 10^{-4}\,M_{\odot}$ exist and form tight binaries with neutron stars, resonance locking may provide a channel to delay or avoid merger that destroys the star.

\vspace{0.3cm}
\noindent \textbf{\textit{Resonance locking of neutron-star g-modes.}}
Neutron stars with chemical stratification display a forest of g-modes with frequencies $\lesssim 200\ \mathrm{Hz}$~\cite{Reisenegger:1992,Kokkotas:1999bd}.  The highest of these frequencies is ultimately set by the geometry of the star and the Brunt-V\"ais\"al\"a frequency, i.e.~the buoyancy frequency at which a slightly displaced fluid element oscillates about its original position in a stably stratified fluid. At some lower frequency, the g-mode forest terminates because the mode periods are longer than the relevant weak reaction rates in the star~\cite{Reisenegger:1992, Haensel:1992zz}, though at which frequency this happens is highly uncertain due to poorly constrained microphysics.  For cold nucleonic stars, this frequency is likely no more than $1\,\rm{Hz}$~\cite{Reisenegger:1992, Lai:1993di}, which is the characteristic value we use here.  

The tidal force of a companion object on a neutron star drives oscillation modes in the latter.   If we define 
\begin{equation}
    \label{eq:mode-driving}
    \sigma = m(\Omega_{\NS} - \Omega_{\orb})\,,
\end{equation}
where $\Omega_{\orb}$ is the orbital frequency, then the resonance condition is ~\cite{Lai:1993di, Lai:2006pr}
\begin{equation}
    \omega_X = \abs{\sigma},
\end{equation}  
for a mode of frequency $\omega_X$. For example, an $\ell=m=2$ g-mode of frequency $\Omega_{g,\ell=m=2}/2\pi \sim 10\ \rm{Hz}$ of a neutron star with spin frequency $\Omega_{\NS}/2\pi \sim 10\ \Hz$ would lock to a companion in a $\Omega_{\orb}/2\pi \sim 5\ \mathrm{Hz}$ orbit. However, the opposite ordering of $\Omega_{\NS}$ and $\Omega_{\orb}$ is possible, in which case angular momentum flows from the orbit to the neutron star.

Generally, tidal modifications to the spin depend sensitively on the rate of dissipation in the deformed body, which for neutron stars is assumed to be small.
For example, \citet{Bildsten:1992my} showed that the collective tidal angular momentum transfer during a typical double neutron star inspiral is insufficient to tidally lock the binary components to one another. They, however, assumed an \textit{equilibrium} tidal model, which, as we have already discussed, is unable to substantially alter the behavior of the binary. We must then consider the dynamics of tides to assess whether their conclusion remains robust in tight compact binaries.

For dynamical tides, the tidal torque is typically \emph{proportional} to the quality factor of the mode, rather than inversely proportional to it~\cite{Zahn:1977mi, Lai:1993di, Burkart:2013}, at least if the inspiral is sufficiently slow to (approximately) saturate the resonance.  For neutron star g-modes that may have quality factors of order $\gtrsim 10^{8}$~\cite{Reisenegger:1992}, the tidal torque could be large and locking is promising.  The main impediment to locking is very low overlaps between neutron star g-modes and the tidal field of the companion. We estimate in the Supplemental Material that neutron star g-modes down to $\sim 10\,\Hz$ may be able to lock solely through linear response. At lower frequencies, locking may still be possible, but it will likely require nonlinear interactions between shorter-wavelength (higher-order) modes~\cite{Weinberg:2012, Weinberg:2013pbi}. Such interactions depend on poorly constrained dense-matter physics. Once a resonance lock is established, however, the resonance condition and angular-momentum balance constrain the coupled spin and orbital evolution. We therefore determine the conditions under which a sustained lock accelerates inspiral or induces outspiral, and assess the observational implications.

We begin by deriving the equations of motion for a binary system with an extended compact body and a much smaller compact body in a leading (nontrivial) order post-Newtonian orbit and in full generality.  We consider the more massive, geometrically larger object to be a neutron star with mass $M_{\NS}$, and the smaller body to be a small (sub-solar mass) black hole with mass $M_{\BH}$, though this fact is not important for this derivation.  The equations of motion for the angular momenta are
\begin{align}
    \label{eq:L-eom-orb}
    \dot L_{\orb} &= \tau_{\tide} +\tau_{\rm GW}; \qquad
    \dot L_{\NS} = -\tau_{\tide},
\end{align}
where $\tau_{\rm GW}$ is the leading-order post-Newtonian gravitational-wave torque, and $\tau_{\tide}$ (which we calculate under various hypotheses below) is the torque of the body on the orbit.  To enforce the resonance locking condition, we convert the angular momenta into angular velocities
\begin{align}
    \label{eq:I-orb-def}
    \Omega_{\orb}  &= L_{\orb}/I_{\orb}; \qquad
    \Omega_{\NS} = L_{\NS}/I_{\NS},
\end{align}
where $I_{\orb}$ is the effective orbital moment of inertia, given, again at leading order and for a quasi-circular orbit, by
\begin{equation}
    \label{eq:Iorb}
    I_{\orb} = M_{\BH}r^2 = M_{\BH}(GM_{\NS})^{2/3}\Omega_{\orb}^{-4/3},
\end{equation}
where $r$ is the orbital separation and we have used Kepler's third law. Equation~\eqref{eq:L-eom-orb} then becomes
\begin{align}
    \dot \Omega_{\orb} &= \frac{\tau_{\tide} + \tau_{\GW}}{I_{\rm orb}} - \Omega_{\orb}\frac{\dot I_{\orb}}{I_{\orb}}; \qquad
    \dot \Omega_{\NS} = \frac{-\tau_{\tide}}{I_{\NS}},
\end{align}
because the neutron star moment of inertia $I_{\NS}$ is constant. 
Differentiating Eq.~\eqref{eq:Iorb} and simplifying, these become (\emph{c.f.} Eq.~(7) of Ref.~\cite{Burkart:2013})
\begin{align}
    \label{eq:Omegadot}
    \dot \Omega_{\orb} &= -3\frac{\tau_{\tide}+\tau_{\GW}}{I_{\orb}}; \qquad
    \dot \Omega_{\NS} = \frac{-\tau_{\rm tide}}{I_{\NS}}.
\end{align}
Interestingly, because of the dependence of the effective orbital moment of inertia on the total angular momentum of the orbit, both $\Omega$'s exhibit the same-sign response to tidal angular momentum transfer. That is, when one increases, so will the other, and vice-versa.  In order to maintain the resonance locking criterion, $\dot \Omega_{\NS} = \dot \Omega_{\orb}$, a particular value of $\tau_{\tide}$ must be chosen:
\begin{equation}
    \label{eq:tidal-torque-unreduced}
    \tau_{\tide} = \frac{\tau_{\GW}}{\quant{I_{\orb}/(3I_{\rm NS}) -1}}.
\end{equation}
Equation~\eqref{eq:tidal-torque-unreduced} implies that without gravitational-wave  torque, given our hypotheses (in particular that the neutron star moment of inertia $I_{\NS}$ and mode frequencies $\omega_X$ do not evolve), no tidal torquing can occur.  Therefore, the inclusion of relativity induces an additional channel for resonance locking not present in the purely Newtonian case.  Physically speaking, this is because gravitational waves serve as a sink for angular momentum; the process would otherwise be kinematically forbidden.  Additionally, as is standard, the tide must be \emph{passive}, so the average rate of energy dissipation inside the neutron star, $\dot Q$ must be positive.  Assuming an insignificant amount of energy is stored in the mode itself, this fixes the sign of $\sigma$ to be the same as $\tau_{\rm tide}$, as
\begin{equation}
    \dot Q = \frac{\sigma}{m}\tau_{\tide}.
\end{equation}

\vspace{0.3cm}
\noindent{}\textbf{\emph{Orbital and spin dynamics during resonance lock.}}
Resonance locking of neutron star g-modes to a small compact object can occur in two different regimes, based on where the resonance lock initiates.  To understand these two regimes, let us substitute Eq.~\eqref{eq:tidal-torque-unreduced} into Eq.~\eqref{eq:Omegadot} to find
\begin{equation}
    \label{eq:inspiral-enhancement}
    \frac{\dot\Omega_{\orb}}{\dot\Omega_{\orb}^{\rm GW}}
    = \frac{I_{\orb}}{I_{\orb}-3I_{\NS}},
\end{equation}
where $\dot\Omega_{\orb}^{\rm GW}=-3\tau_{\GW}/I_{\orb}$ is the frequency evolution due to gravitational-wave emission alone, evaluated at the same orbital separation. Therefore, the two regimes depend on whether locking is established when $I_{\orb}>3I_{\NS}$ or when $I_{\orb}< 3I_{\NS}$.

If the resonance lock is initiated with $I_{\orb}>3I_{\NS}$, then ${\dot\Omega_{\orb}}/{\dot\Omega_{\orb}^{\rm GW}} > 1$, and the inspiral is accelerated. Physically, the tidal torque transfers angular momentum from the orbit to the neutron star, so this configuration requires $\Omega_{\orb}-\Omega_{\NS} > 0$. As a result, the neutron star spins up because its spin angular momentum increases, and the small black-hole companion inspirals faster, because the orbital angular momentum decreases faster than it would under gravitational-wave emission alone. As the body inspirals, the effective orbital moment of inertia decreases with the orbital separation, while the neutron star moment of inertia remains constant. As $I_{\orb}$ approaches $3I_{\NS}$ from above, the tidal torque required to maintain the resonance lock grows without bound in this description. Before this though, the description of the system we provide breaks down.

If the resonance lock is initiated with $I_{\orb}<3I_{\NS}$, then $\dot\Omega_{\orb}<0$, and the companion begins to outspiral. Physically, the tidal torque feeds angular momentum from the faster-spinning neutron star to the orbit at a rate greater than that removed by gravitational-wave emission.  This configuration requires  $\Omega_{\NS}>\Omega_{\orb}$.  As the companion outspirals, the effective orbital moment of inertia increases with the orbital separation, while the neutron star moment of inertia remains constant. As $I_{\orb}$ approaches $3I_{\NS}$ from below, the tidal torque required to maintain the resonance lock again grows without bound, though the model surely breaks down before this occurs.

\begin{figure*}
    \centering
        \includegraphics[width=0.49\textwidth]{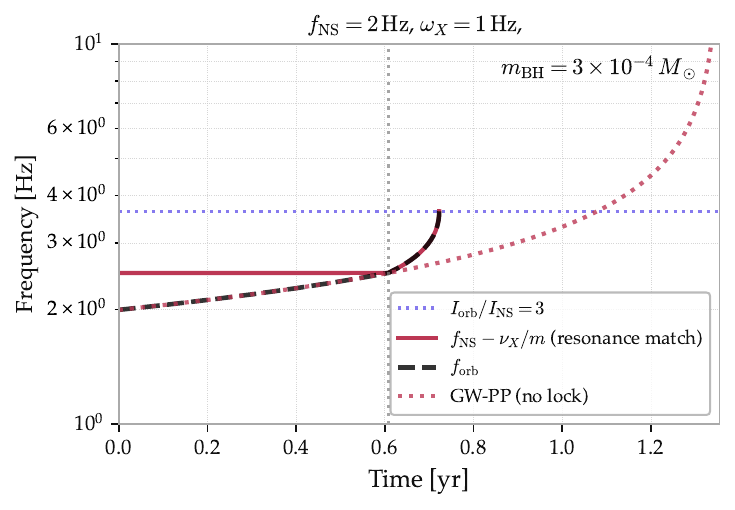}
    \includegraphics[width=0.49\textwidth]{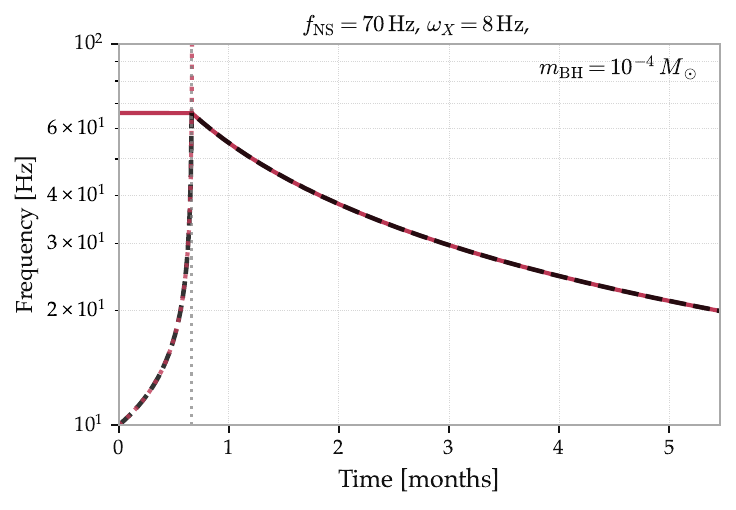}    
    \caption{Examples of the frequency evolution of binaries that undergo locking. 
    Red solid lines show the frequency which must match the orbital frequency for the resonance locking condition to be satisfied. 
    Orange dotted lines show the gravitational-wave-driven evolution of the orbital frequency in the point-particle model.  
    Black dashed lines show the full evolution of the orbital frequency, including its evolution when approaching resonance lock. 
    The vertical dotted gray line shows the time at which locking begins because Eq.~\eqref{eq:mode-driving} is satisfied, while the horizontal dotted purple line marks the frequency pole of the tidal torque.
    In all evolutions here, the neutron star has mass $1.4\,M_{\odot}$. On the left panel, observe that the system locks when $I_{\orb } > 3I_{\NS}$, and so the inspiral is accelerated, while on the right panel, lock occurs with $I_{\orb } < 3I_{\NS}$, so the inspiral is reversed and the binary outspirals. 
    } 
    \label{fig:dynamics}
\end{figure*}

Figure~\ref{fig:dynamics} illustrates the two regimes of resonance-locked orbital evolution. 
In the left panel, locking begins when $I_{\orb}>3I_{\NS}$, and the orbital frequency increases more rapidly than predicted from gravitational-wave emission alone, accelerating the inspiral. In the right panel, locking begins when $I_{\orb}<3I_{\NS}$, and the orbital frequency decreases despite continued gravitational-wave emission, so the binary outspirals.  In the second case, the spin of the NS serves as the reservoir of energy for the system, so  the total gravitational-wave energy emitted over the $\sim \mathrm{years}$-long inspiral-outspiral is much larger than the initial orbital energy of the primordial black hole.

\vspace{0.3cm}

\noindent{}\textbf{\emph{Astrophysical implications.}}
Our calculations apply to any binary composed of a geometrically large, gravitationally compact object and a much lower mass companion, but the prototypical case is a neutron star with a subsolar-mass primordial black hole.  The abundance of subsolar mass black holes is constrained by dark matter searches, which imply that black holes in the mass range $\sim (10^{-6}$--$10^{-2})\,M_{\odot}$ can represent no more than $\lesssim 1 \%$ of dark matter~\cite{EROS-2:2006ryy, Niikura:2017zjd,Garcia-Bellido:2017knh,Carr:2026hot}.  Assuming a typical dark matter density of $\sim 0.5\,\rm{GeV/cm^3}$ ($\approx 0.013\,M_{\odot}/\rm{pc}^3$) and a primordial black hole fraction of $1\%$, this leaves $\lesssim 130$ such black holes per cubic parsec. At lower masses, a larger fraction of dark matter may be black holes, and a larger number of black holes are present per unit mass; therefore, the number density of lower-mass primordial black holes is possibly much larger.  Per the arguments above, lower mass black holes are likely to outspiral if resonance locked to neutron star g-modes.  Therefore close capture scenarios may not necessarily lead to neutron star destruction~\cite{Baumgarte:2024iby, Baumgarte:2024ouj} if resonance locking is generic for these systems, as we discuss further in the End Matter.

Another astrophysical implication of resonance locking applies to pulsar binaries.  Observationally, if the neutron star in question is a pulsar, the presence of a $10^{-3}\,M_{\odot}$ companion should be quite evident, unless the orbit is very tight. Low-mass companions with masses $\sim 10^{-8}$--$10^{-3}\,M_{\odot}$ have been detected around pulsars~\cite{Wolszczan:1992zg,Konacki:2003xa,Vleeschower:2024nbp}, demonstrating that neutron stars can host companions with very small mass ratios. Dynamical exchange has been proposed to explain the planetary companion orbiting the neutron star--white dwarf binary PSR B1620-26~\cite{Sigurdsson:2003is}.
While most pulsar planets have long orbital periods ($\gtrsim 1\,{\rm day}$), at least one has been observed with a mass $\sim 10^{-3}\,M_{\odot}$ and a period of $\sim 2.2\,{\rm h}$~\cite{Bailes:2011}. Detecting a planetary-mass companion to a pulsar with an orbital period $\lesssim 1\,{\rm s}$ would demonstrate the existence of an exceptionally dense low-mass object, since an ordinary planet would be tidally disrupted in such an orbit. 
However,  the impact of such a planet on the pulse profile of a $\sim \Hz$ spin-frequency pulsar may be complicated.  

Substantially lower-mass objects, $\lesssim 10^{-10}\,M_{\odot}$, and objects in very tight orbits, may lie unresolved by pulsar timing~\cite{Nitu:2022ath}. The NANOGrav 15-year dataset contains 68 well-timed millisecond pulsars, with timing precision of $\sim 1\,\mu \rm{s}$~\cite{NANOGrav:2023hde}.  Thus, while detectability depends on the pulsar's timing noise, the orbital inclination, and the observing baseline,  pulsars in binaries with center-of-mass-to-barycenter distances of about $0.1 \,$km (provided their period is $\lesssim 1\,\textrm{yr}$) should have visible modifications to the pulsar times of arrival. For a $10^{-9}\,M_{\odot}$ companion to produce a timing amplitude of $\sim1\,\mu{\rm s}$, its orbital separation would need to be $\sim4\times10^8\,{\rm km}$.  This corresponds to an orbital period of $\sim4\,{\rm yr}$ for an edge-on orbit around a $1.4\,M_{\odot}$ neutron star, and thus such systems are not interesting targets for the work discussed here, barring substantial orbital hardening.

\begin{figure}
    \centering
    \includegraphics[width=0.99\linewidth]{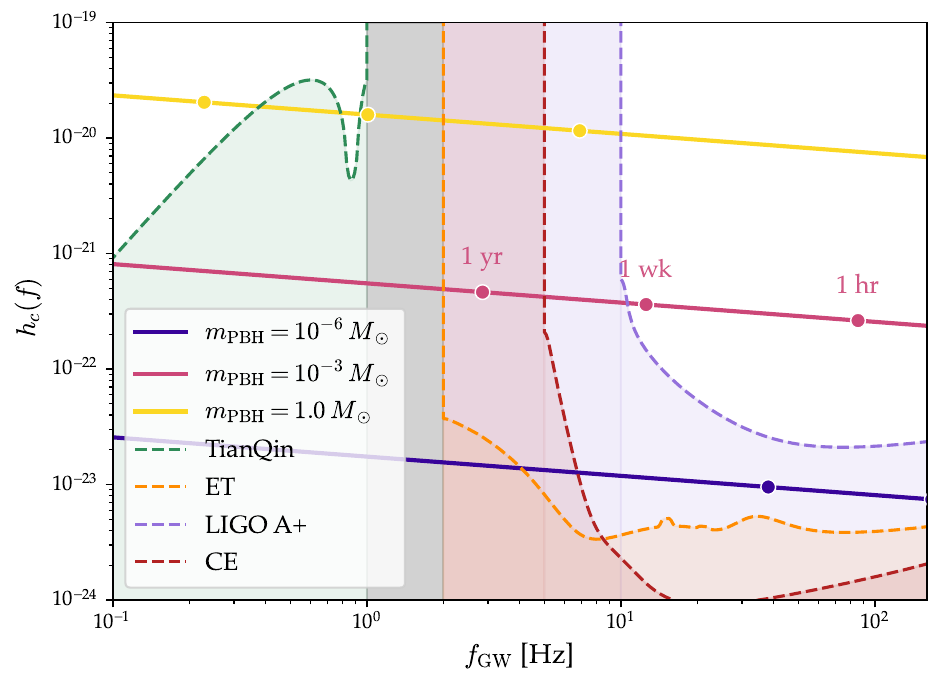}
    \caption{Example characteristic strain of a $1.4\,M_{\odot}$ neutron star merging with a primordial black hole laid over sensitivity curves for various detectors; the source is at $10\,\mathrm{Mpc}$.  We show results for 3 different primordial black hole masses in colored solid lines, and for 4 detectors at proposed sensitivity in dashed lines.  We use {\sc PyCBC}~\cite{pycbc} to retrieve estimates of the power spectral densities for various interferometer configurations  (TianQin, green dashed~\cite{TianQin:2015yph}; Einstein Telescope (ET), orange dashed~\cite{Hild:2010id, Evans:P1600143}; LIGO at A+ sensitivity, purple dashed~\cite{Barsotti:T1800042}; Cosmic Explorer, red dashed~\cite{LIGOScientific:2016wof, Evans:P1600143}). Colored shaded regions mark the regions nominally unseen by the respective ground-based detectors. Dots mark  1 hour, 1 week, and 1 year to merger for the given binary. A gray vertical band marks the frequency region with no proposed detector coverage separating space and ground-based detectors. }
    \label{fig:gw-strain}
\end{figure}

A final astrophysical implication of the existence of primordial black hole-neutron star inspirals is that they would be natural targets for ground (and possibly space-based) gravitational-wave detectors. 
Figure~\ref{fig:gw-strain} displays the characteristic strain predicted for binaries evolving solely through gravitational-wave emission, together with the sensitivity curves of current and future detectors. These estimates provide a reference for detectability; the effect of resonance locking on the SNR depends on the resulting frequency evolution and observing duration.  However, Fig.~\ref{fig:dynamics} implies that typically, the SNR of a resonance locked system will be comparable to, or in the outspiral case, possibly much larger than, the corresponding point-particle system.

Very broadly, we find that $10^{-4}\,M_{\odot}$ primordial black holes will not be detectable by conventional standards at A+ sensitivity and a distance of $10\,\rm{Mpc}$ ($\rm{SNR}\sim 5$), whereas $10^{-3}\, M_{\odot}$ primordial black hole-neutron star binaries should be detectable in principle ($\mathrm {SNR}\sim 16$ at 10 Mpc). Sources with $M_{\rm BH}>10^{-6}\,M_{\odot}$ should be detectable in Cosmic Explorer, having $\mathrm{SNR}>10$ at $10\, \mathrm{Mpc}$. Importantly, the signal-to-noise ratio of these systems scales comparatively weakly with mass ratio, $\mathrm{SNR}\propto \eta^{1/2}$. 

Nonlinear resonance locking has also been proposed in comparable-mass neutron-star binaries, where it may substantially affect the gravitational-wave phase while the binary continues to inspiral~\cite{Kwon:2024zyg,Kwon:2025zbc}.  In the intermediate mass ratio regime, both nonlinear modification and secular changes to neutron star rotation may be important. Assessing the implications for current ground-based searches for subsolar-mass black holes~\cite{LVK:2022ydq,Kacanja:2026byy, LIGOScientific:2026kef, LIGOScientific:2026wxz} requires gravitational-wave models that include the locked evolution. Further derivations of the conditions for establishing and breaking a resonance lock, and of the effective tidal dissipation during the lock, are provided in the Supplemental Material~\cite{HegadeKR:2024agt,Ho:1998hq,Ripley:2023lsq,Ripley:2023qxo,Yu:2016ltf}.

Our results show that neutron star--primordial black hole binaries can exhibit orbital evolution strikingly different from that of typical compact binaries. Once a sustained resonance lock is established, dynamical tides can accelerate inspiral or drive outspiral despite continued gravitational-wave emission. In the latter case, the expanding orbit draws on the neutron star's spin angular momentum, analogous to the outward migration of moons around rapidly rotating giant planets. These changes could leave distinctive signatures in pulsar timing and gravitational waves. Dynamical tides therefore merit particular attention in neutron-star binaries with large mass asymmetries, both when searching for these sources and when using them to constrain primordial black hole dark matter.

\vspace{0.3cm}
\begin{acknowledgments}
\noindent{}\textbf{\textit{Acknowledgments}.}
The authors acknowledge support from the Simons Foundation through Award No. 896696, the Simons Foundation International through Award No. SFI-MPS-BH-00012593-01, the NSF through Grants No. PHY-2207650 and PHY-25-12423, and NASA through Grant
No. 80NSSC22K0806. This work was also supported
by NSF within the framework of the MUSES collaboration, under grant number OAC-2103680. 
\end{acknowledgments}

\vspace{0.3cm}
\noindent{}\textbf{\textit{Data Availability.}}
Plotting scripts used in this Letter will be available on GitHub upon publication.

\vspace{0.3cm}
\noindent{}\textbf{\textit{AI statement.}}
Claude Sonnet 4.6 and Claude Opus 5, as well as Codex GPT-6.1 Sol Ultra, were used to generate figures and verify the correctness of the draft. Figure scripts were verified by I.L.  The resonance locking breaking formula was originally calculated by AI, and verified by the authors in various ways, including using computer algebra software.

\vspace{0.3cm}
\noindent{}\textbf{\textit{End Matter.}}
Gravitational-wave emission during a close encounter provides one possible formation channel for neutron star--primordial black hole binaries~\cite{Peters:1963ux,Quinlan:1987}. In the point-particle approximation, black holes with asymptotic relative speed $v_{\rm rel}$ become captured if their periastron distance is less than
\begin{align}
&r_{p,\max}\simeq
\left(\frac{85\pi}{6\sqrt{2}}\right)^{2/7}
\frac{G\,\mu^{2/7}M_{\NS}^{5/7}}
{c^{10/7}v_{\rm rel}^{4/7}}\,,
\\ \nonumber
&\simeq 36.2\,{\rm km}
\left(\frac{\mu}{10^{-3}M_\odot}\right)^{2/7}
\!\!\left(\frac{M_{\NS}}{1.4M_\odot}\right)^{5/7}
\!\!\left(\frac{v_{\rm rel}}{300\,{\rm km/s}}\right)^{-4/7}\!\!\!\!,
\end{align}
where $\mu=M_{\BH}M_{\NS}/(M_{\BH}+M_{\NS})\approx M_{\BH}$ is the reduced mass, and we use a fiducial relative speed of $300\,{\rm km\,s^{-1}}$ for the numerical normalization~\cite{Genolini:2020ejw}. The maximum periastron decreases with the primordial black hole mass and increases as the encounter speed decreases. For sufficiently small masses, capture outside the neutron star therefore requires slower encounters, which could be realized through many-body interactions. For passages near the stellar surface, dynamical tides may also contribute by transferring orbital energy into neutron star oscillation modes~\cite{Genolini:2020ejw}.  While estimates for these effects exist in the literature~\cite{Press:1977,Capela:2013yf,Defillon:2014wla,Pani:2014rca}, we believe a realistic encounter between a neutron star and primordial black hole should be governed essentially by nonlinear effects in the hydrodynamic flow of the neutron star. Multi-body encounters with other stars can also exchange orbital energy and angular momentum, modifying both capture and the subsequent evolution of the binary, and may be the dominant source of primordia capture~\cite{Gottlieb:2026kpg}. A quantitative formation model must therefore assess these processes together with the primordial black hole abundance and encounter velocity distribution. In general, it is more challenging to capture a lower mass black hole onto a neutron star, either by tidal effects or gravitational-wave emission. However, at a fixed dark-matter mass fraction, lower mass black holes are more abundant, and therefore the odds of a capture of \emph{any} primordial black hole onto a given neutron star are higher.  

How is this scenario affected by the possibility of resonance locking? If the PBH becomes resonance locked, with $3I_{\NS}>I_{\orb}$, which is generic for low-mass primordial black holes, then the system will instead outspiral toward the orbit where $3I_{\NS} = I_{\orb}$. For a primordial black hole with $M_{\rm BH}\sim 10^{-8}\, M_{\odot}$, this resonance-lock-breaking radius is 
$\gtrsim  5\times10^4\,({GM_{\NS}}/{c^2}) \simeq 10^5\,{\rm km} \sim 10^{-3}\,{\rm AU}$, at which point gravitational inspiral times would exceed a Hubble time. Therefore, resonance locking may drive the primordial black hole far enough away from the neutron star to prevent coalescence. For higher mass primordial black holes ($\sim 10^{-4}\, M_{\odot}$), the inspiral time from the resonance-lock breaking radius is only $\sim \rm{years}$, and so these objects may very well merge with their host neutron stars eventually, but the predicted rate of such mergers is constrained to be very small.

If a $10^{-4}\,M_{\odot}$ primordial black hole did become resonance locked to a neutron star g-mode in a pulsar, close to the critical frequency, where $3I_{\NS}=I_{\orb}$, then it would likely manifest as a sudden and apparently non-smooth transition to a much larger rate of period evolution.  In particular, there would be a change in the braking index of the pulsar~\cite{Manchester:2005, Espinoza:2011}
\begin{equation}
    n \equiv \frac{\ddot ff}{\dot f^2},
\end{equation}
which during a lock (if other sources of braking are unimportant) becomes
\begin{equation}
    \label{eq:braking-index-locked}
    n = \frac{11 I_{\orb} - 21 I_{\NS}}{3 (I_{\orb } - 3 I_{\NS})} \frac  {\Omega_{\NS}} { \Omega_{\orb}}.
\end{equation}
Therefore, the braking index depends on the particular orbital configuration when the lock is achieved.  If the configuration is such that $I_{\orb}\gg3I_{\NS}$, then the braking index will become $11/3 \frac{\Omega_{\NS}}{\Omega_{\orb}}$, which is less than $11/3$ by passivity.  Nonetheless, the pulsar will be spinning up rather than spinning down, which is observed only in a handful of cases tied to acceleration of pulsars along the line-of-sight within globular clusters~\cite{Anderson:1990, Phinney:1992}. It is also possible that a very small companion becomes resonance locked, in which case electromagnetic-driven spindown can become important, in addition to the tidal torque.  This may serve to drive the companion to larger separations and lower frequencies, even if $I_{\orb}\gg3I_{\NS}$.  If $I_{\orb}< 3I_{\NS}$, on the other hand, and if $11I_{\orb} > 21 I_{\NS}$, then the pulsar will spin down but the braking index will be negative. Otherwise, if $I_{\orb}< 3I_{\NS}$, the pulsar will spin down with a positive braking index.

Finally, it may be the case that the process of resonance locking is not a clean monotonic transition in the detuning frequency, and even after the lock, the frequency executes cycles around the implied locking ``fixed-point".  For example, in Ref.~\cite{Burkart:2013jua}, a large number of possible locking phenomenologies are observed. In this case, the implied braking frequency of the pulsar could be transiently quite different from the one implied by Eq.~\eqref{eq:braking-index-locked}.  Complex behavior in the detuning frequency may occur on multiples of the damping timescale $1/\gamma$, with $\omega/\gamma$ the mode quality factor. Since mode quality factors may be $\gtrsim 10^8$ for $1\,\mathrm{Hz}$ modes, the damping timescale may be $\sim \mathrm{years}$ for neutron stars. Therefore, extended observations of a large number of Galactic pulsars for long periods of time are likely the only way to detect such fluctuations.

\bibliography{references}
\clearpage
%% Body of the Supplemental Material.  No preamble, no \begin{document}:
%% included by main.tex (appended to the Letter) and by
%% supplement-standalone.tex (separate file for the journal submission).

\setcounter{equation}{0}\setcounter{figure}{0}
\setcounter{table}{0}\setcounter{section}{0}
\renewcommand{\theequation}{S\arabic{equation}}
\renewcommand{\thefigure}{S\arabic{figure}}
\renewcommand{\thetable}{S\arabic{table}}

\section{Supplemental Material}
\subsection{Essential Physics}

In the absence of tidal effects, the inspiral of two compact objects proceeds due to gravitational waves~\cite{Peters:1963ux}
\begin{equation}
\dot E \approx \dot E_{\GW} \overset{(\text{LO})}{=} -\frac{32c^5}{5G}\eta^2 x^5f(e),
\end{equation}
where $\eta= {q}/{(1+ q)^2}$ is the symmetric mass ratio, with $q = M_{\BH}/M_{\NS}$ the mass ratio (and the overset LO indicates the expression holds at leading order in the post-Newtonian expansion).  Additionally, Kepler's third law tells us that
\begin{equation}
   x \equiv \quant{\frac{GM \Omega_{\orb}}{c^3} }^{2/3}\overset{(\mathrm{LO})}{=}\frac{GM}{ac^2},
   \label{eq:x-PN}
\end{equation}
with $\Omega_{\orb}$ the orbital frequency of the binary, $M=M_{\BH}+M_{\NS}\approx M_{\NS}$ is the total mass of the binary.  The function $f(e)$ depends only on the eccentricity and is an enhancement factor ($f(e) \geq 1$) to the radiated power. There is no reason to expect that the longer-period binaries we look at here should have low eccentricity $e\sim 0$, but unless otherwise stated we take $e=0$ because the tidal analysis is more complicated without this assumption.  We do not expect the core conclusions to be changed by the inclusion of eccentricity, unless the eccentricity is excessively large.  

The orbital energy of the system is 
\begin{equation}
    E_{\orb} = -\frac{GM_{\NS}M_{\BH}}{2a},
\end{equation}
where $a$ is the semi-major axis (or the constant radius of the orbit, with our zero-eccentricity assumption).  If the symmetric mass ratio is sufficiently small, then the inspiral timescale,
\begin{equation}
    \label{eq:gw-timescale}
    \frac{t^{(\GW)}_{\rm inspiral}}{GM/c^3} \sim \frac{E_{\orb}} {\dot E_{\GW}}  = \frac{5c^8}{64G^4\eta}\frac{a^4}{ M^4 },
\end{equation}
is exceptionally long, even if $x$ is not small (equivalently, even if $a\sim  \mathrm{tens\ of}\ M$).  

Neutron stars have another important timescale, the dynamical timescale:
\begin{equation}
    t_{\rm dyn} \equiv \frac{1}{\sqrt{G\rhoc}}\sim \sqrt{\frac{R_{\NS} c^2}{GM}} \frac{R_{\NS}}{c}.
\end{equation}
Interestingly, the dynamical timescale of neutron stars is not too different from the gravitational timescale $t_{\rm grav} \equiv GM/c^3$, since the compactness $C_{\NS} \equiv GM_{\NS}/(R_{\NS}c^2) \approx 0.2$ for typical neutron stars.

Oscillation modes of a \textit{nonrotating}, fluid star without stratification will be spaced fairly evenly in frequencies $\omega_{\rm {fund}} \gtrsim 2\pi/t_{\rm dyn}$.  Therefore, these modes are unlikely to be resonantly excited until $\Omega_{\orb}\sim \omega_{\rm fund}$, which requires $x = (2 \pi)^{2/3} C_{\NS}$, which, for a typical neutron stars means $x \sim 0.7$, deep in the highly relativistic, dynamical regime.  However, if the star is rotating, or if the star is stratified, or both, then there will be a spectrum of inertial modes, g-modes, or both.  These modes have frequencies related to the rotation period or the Brunt-V\"ais\"al\"a frequency respectively, which in most cases, will be small compared to the fundamental frequency. This is what allows a neutron star mode to become resonantly locked. 

In the inspiral of non-rotating neutron stars with comparable masses, orbital energy is transferred into (1) the physical deformations of the neutron star and (2) the gravitational field via tidal interactions.  It is generally assumed that the most efficient energy transfer is mediated by the ``slow" driving of the non-radial, fundamental $\ell=2$ mode of the star, which is called the \emph{equilibrium tide}.  In this case, the radiation-reaction timescale is very short owing to the relatively large symmetric mass ratio ($\eta \sim 1/4$), so modes are not driven near their resonant frequencies for very long, as the frequency quickly chirps through the intrinsic oscillation frequency.  Therefore, the system is dominated by the very large response of the equilibrium tide, and resonances are subdominant.  

A consequence of this reasoning is the very strong scaling of the tidal response with the orbital separation.  In particular, the amplitude of the equilibrium tidal mode (with frequency $\omega_{\ell mn} = \omega_{220} = \omega_{20}$, since the star is nonrotating) is proportional to
\begin{equation}
    a_{20}(\omega) = \frac{1}{\omega^2_{20} - \omega^2}\frac{Q_{20}}{A_{20}^2} d_{20}(\omega)    
\end{equation}
where $Q_{20}$ is the tidal overlap, which is a purely geometric factor dependent on the mode profile, $A_{20}$ is a normalization factor, and $d_{20}\propto M_{\BH}/r^3$ is the tidal field induced by the black hole on the neutron star.  The frequency $\omega$ is not close to $\omega_{20} \sim 2\pi/t_{\rm dyn}$ unless the orbit of the point particle has radius $r \sim R_{\NS}$, so the mode amplitude is suppressed at low orbital frequency (large orbital separation) by both the tidal tensor and the dynamical response.  

On the other hand, $g$-modes and inertial modes will also have mode amplitudes 
\begin{equation}
    a_{X}(\omega) = \frac{1}{\omega^2_{X} - \omega^2}\frac{Q_{X}}{A_{X}^2} d_{\ell}(\omega),
\end{equation}
but may be resonantly excited at much lower frequencies, and thus, g-modes are not penalized as strongly by large orbital separation.  In typical analyses of comparable-mass neutron star inspirals, the usual g-modes and inertial modes are neglected because of the small overlaps $Q_{X}$ between these modes and the tidal fields.  However, fundamentally, the unimportance of these modes has more to do with the details of the driving mechanism.  If the quality factor of the orbit is small, so that $\dot a/a$ is large, then, in frequency space, the driving force of the binary is delocalized. 

To show this, let us consider a simplified model of the driving tidal force due to the companion 
\begin{equation}
    d_\ell(\omega) \propto \int_{-T}^T d_{\ell}(t)\exp(-i\omega t)dt,
\end{equation}
where $T$ is some timescale, which for our case we can imagine to be the time range where the orbit is close to resonance. 
The time dependent driving force of the neutron star tide due to the black hole will look like 
\begin{equation}
    d_{2}(t) \propto\frac{M_{\BH}}{r^3(t)}e^{i\int \Omega_{\orb}(t)dt},   
\end{equation}
so $d_{2}(\omega)$ is given, in the stationary phase approximation, by
\begin{equation}
    \label{eq:tidal-driving-force-fd}
    d_2(\omega)\approx\frac{M_{\BH}}{r^3(t_{\omega})}\frac{\sqrt{2\pi}}{\sqrt{\dot\Omega_{\orb}(t_{\omega})}}\exp\quantb{i\Phi(t_{\omega}) + i\frac{\pi}{4}},
\end{equation}
assuming $\dot \Omega(t) > 0$ and taking $\Phi(t) = \int_{-T}^t \Omega_{\orb}(t) dt$, with $t_{\omega}$ the stationary time at which $\Omega_{\orb}(t_{\omega})= \omega$.  If $\dot \Omega$ is large, then the driving force is suppressed in the frequency domain.  Conversely though, if $\dot \Omega$ is small, the force is amplified.
This is the case for the post-Newtonian decay of binary orbits with small mass ratio.  In particular, assuming no back reaction of the tide on the orbit, and no fluid dissipation, the maximum mode amplitude can be calculated according to \cite{Lai:1993di}, 
\begin{equation}
    |a_{\alpha,\max}| \propto \frac{GM_{\BH} Q_\alpha}{a^{\ell_{\alpha}+1}}\left(\frac{\pi}{\dot\Omega_{\rm orb}}\right)^{1/2}_{\rm res}.
\end{equation}

At some point in the orbit, fluid dissipation or nonlinear interactions can no longer be ignored. This will occur when the quality factor of the mode is comparable to $\Omega^2/\dot \Omega$. For neutron star g-modes, the quality factor may be $\sim 10^{8}$ (though this is highly uncertain, as we discuss below).  For generic values of mode frequencies and neutron star spins, these effects will become relevant around masses of $10^{-4}\text{--}10^{-3}\,M_{\odot}$.  However, it is important to recognize that $\omega_X=m(\Omega_{\NS}-\Omega_{\orb})$ can have solutions with $\omega_X\sim 1\,\mathrm{Hz}$, $\Omega_{\NS}\sim 1\,\mathrm{Hz}$, $\Omega_{\orb}\ll\mathrm{1\,Hz}$.  In this case the inspiral is arbitrarily slow, and the resonance will almost certainly saturate.  In this regime, further, the smallness of the black hole mass and the largeness of the orbital separation are less important than the detuning frequency.  This is the regime of interest in this paper.

\subsection{Resonance Locking and Saturation}
We first discuss the detailed  physics by which resonance locking may occur, because similar concepts will be involved in describing the saturation of the resonance at $I_{\orb}/I_{\NS} \approx  3$.  We first describe the case of an object inspiraling from spatial infinity, because we view it as the most astrophysically relevant.  In the absence of any substantial tidal effects, the system evolves under gravitational-wave damping alone: 
\begin{align}
    \dot \Omega_{\orb} &= \frac{-3\tau_{\GW}}{I_{\orb}};\\
    \dot \Omega_{\NS} &= 0.
\end{align}
In practice, neutron stars are expected to have magnetic fields, and thus, they will spin down through magnetic braking as well. We discuss this briefly below.

If the resonance is sufficiently sharp, i.e.\ the mode width $\gamma_\alpha$ is narrow compared to the chirp rate, and if the mass ratio is small enough, then a lock will be achieved. Whether a lock can be established is ultimately determined by the efficiency of internal dissipation. This is quantified in the literature using a dimensionless number called the tidal quality factor~\cite{Fuller:2016}:
\begin{equation}
    Q_{\rm tide} \equiv 3k_2 \frac{M_{\BH}}{M_{\NS}}\frac{R_{\NS}^5}{r^5}\Omega_{\orb}t_{\tide},
    \label{eq:Qtide}
\end{equation}
where $t_{\tide}$ is the time scale associated with tidal orbital energy being dissipated via tides, and $k_2$ is the static tidal Love number\footnote{We use the static Love number to normalize the effective tidal quality factor. This choice does not neglect dynamical tides, whose effects enter through the physical tidal timescale $t_{\tide}$. Changing the Love-number normalization rescales the quoted quality factor without changing the tidal torque, provided the response is treated consistently.}. A large tidal quality factor corresponds to a long tidal evolution timescale. Near a mode resonance, however, dynamical tides can enhance the tidal torque and shorten $t_{\tide}$, even when internal dissipation is weak. A necessary condition for resonance locking is that the available tidal torque can sustain the angular-momentum transfer required by \mainEqTorque. Establishing a lock also requires that the coupled evolution can reach and maintain this state.

Earlier calculations found insufficient tidal angular-momentum transfer to synchronize comparable-mass neutron-star binaries~\cite{Bildsten:1992my}, and high-order g-modes have small overlaps with the companion's tidal field~\cite{Lai:1993di,Reisenegger:1992}. These results do not, however, exclude resonance locking in the extreme-mass-ratio systems considered here. A much lower-mass companion slows the gravitational-wave-driven evolution, allowing more time for a resonant response to develop, while weak mode damping can enhance the torque near resonance. We therefore compare next the attainable resonant tidal torque with the torque required to maintain a lock.

Let us then analyze this problem quantitatively, and consider the various channels that may allow a resonance lock to occur. To do this, we compare the tidal timescale $t_{\tide}$ to the inspiral timescale $t^{(\GW)}_{\rm inspiral}$ from Eq.~\eqref{eq:gw-timescale}. We first rewrite Eq.~\eqref{eq:gw-timescale} as
\begin{equation}
     \Omega_{\rm orb}t^{(\GW)}_{\rm inspiral} =\frac{5}{64}\frac{1}{\eta}\quant{\frac{c^2r}{GM}}^{5/2}
 \end{equation}
and Eq.~\eqref{eq:Qtide} as
\begin{equation}
    \Omega_{\orb}t_{\tide} = Q_{\rm tide} \times \frac{  M _{\NS}r^5}{3k_2 M_{\BH} R_{\NS}^5}.
\end{equation}

We now need plausible estimates for the value of $Q_{\tide}$. From~\cite{Ripley:2023qxo}, we rewrite the above expression in terms of the dimensionless dissipative tidal deformability $\Xi$ (using e.g.~Eq.~(24) of~\cite{Ripley:2023qxo}, for a circular orbit of only one extended object, taking $t_{\tide} = \abs{E_{\orb}/\mathcal F_{\rm diss}}$),
\begin{equation}
    t_{\tide} = \frac{c^{13}r^5}{18G^6} \frac{1}{\Xi M_{\BH} M_{\NS}^5\Omega_{\orb}^2}.
\end{equation}
This gives 
\begin{equation}
    \Omega_{\orb}t_{\tide} = \overbrace{\quantb{\frac{\Lambda}{16 \Xi }\frac{c^3}{GM_{\NS}\Omega_{\orb}}}}^{Q_{\tide}}\times \frac{r^5M_{\NS}}{3 k_2M_{\BH} R^5_{\NS}},
\end{equation}
where 
\begin{equation}
    \Lambda \equiv \frac{2}{3} k_2 \quant{\frac{c^{2}R_{\NS}}{GM_{\NS}}}^5
\end{equation}
is the tidal deformability. 

We have therefore developed an interpretation of the tidal quality factor as a measure of the conservative to dissipative tidal deformability, modulated by the dimensionless frequency. Constraints based on GW170817, and nuclear theory indicate that $\Lambda/\Xi$ is most likely at least 1, and potentially much greater than 1~\cite{Ripley:2023lsq}. This indicates that for the orbits we care about, the tidal quality factor \emph{from adiabatic tidal dissipation alone} should be at least $\sim 10^3$ (though perhaps larger, e.g. $10^6-10^9$).  From these estimates, we can then coarsely estimate 
\begin{equation}
    \Omega_{\orb} t_{\tide} \sim 10^3\times \frac{1}{\eta} \times \frac{r^5}{R_{\NS}^5} \sim 10^6 \times \quant{\frac{2\, \rm{kHz}}{\Omega_{\orb}}}^{10/3}, 
\end{equation}
which, for our problem, is $\sim 10^{16}$. This tidal time scale is far too large to be important, which establishes that for such low frequency orbits, only \emph{dynamical} tides can be relevant.

For dynamical tides, the above arguments are different because the tidal torque is dependent on the details of the mode being locked to. Following~\cite{Burkart:2013}, we estimate the tidal torque as
\begin{equation}
    \tau_{\tide} = \frac{12 \pi}{5} E_{\orb} \frac{M_{\BH}}{M_{\NS}}\quant{\frac{R_{\NS}}{r}}^5Q_{\alpha}^2\quant{\frac{\omega_{\alpha} \gamma_{\alpha}}{\delta\omega_{\alpha}^2 + \gamma_{\alpha}^2}}
    \label{eq:tautide-for-dyn-tides}
 \end{equation}
 where $\omega_{\alpha}$ is the mode frequency, $\gamma_\alpha$ is the mode damping rate, $Q_{\alpha}$ is the tidal overlap of the mode, and we have taken $\ell=m=2$ for this estimate.  We assume throughout this work that the mass ratio is sufficiently small and the internal sound speed is sufficiently large, so that excited resonant modes can be treated as standing waves~\cite{Burkart:2013}.

Generally, tidal overlaps of high-order g-modes are expected to be small, which is primarily because these modes have many nodes and therefore are rapidly oscillating radially.  For example, Ref.~\cite{Yu:2016ltf} gives $Q_a\sim 3.5\e{-3}n_a^{-5/2}$ with $n_a$ the number of radial nodes.  However, nonlinear combinations of modes may combine to form effective modes with much longer wavelengths, increasing the tidal overlap~\cite{Weinberg:2012}.  As it stands, computing the nonlinear oscillation spectrum of a neutron star at low frequency is a challenging task, in part because of the density of g-modes at low frequency and the incredibly small damping times of high-order g-modes.  Furthermore, the equation of state of neutron stars is unknown, and even the relevant species which determine the compositional gradients (e.g.~nucleons or hyperons) is unknown. We instead give the necessary condition for a resonance lock to occur, and leave to future work the task of determining which neutron star models are consistent with such conditions.

Evaluating Eq.~\eqref{eq:tautide-for-dyn-tides} for the torques induced through dynamical tides at exact resonance, $\delta\omega_\alpha=0$, and using $M_{\BH}/M_{\NS}\simeq\eta \ll 1$, gives the maximum tidal torque magnitude in the steady linear response:
\begin{align}
    \frac{|\tau_{\tide}|}{|E_{\orb}|} \sim  8\e{-6}\frac{\eta}{10^{-3}}\quant{\frac{10 R_{\NS}}{r}}^5 \frac{Q_{\alpha}^2}{(1\e{-3})^2} \frac{\omega_{\alpha}/\gamma_{\alpha}}{10^{8}}.
    \label{eq:estimate-tau-tide-dyn}
\end{align}
Before comparing the available and required torques, let us discuss the damping assumed in the numerical normalization of Eq.~\eqref{eq:estimate-tau-tide-dyn}. We used here the illustrative value $\omega_\alpha/\gamma_\alpha=10^8$, rather than a damping rate calculated for a specific neutron-star model. Electron shear viscosity contributes to the damping~\cite{Cutler:1987}, with shorter-wavelength modes generally more strongly affected, while the rate depends on the mode eigenfunction, stellar temperature, superfluidity, and crustal physics~\cite{Reisenegger:1992}. Nonlinear mode interactions may introduce additional dissipation~\cite{Weinberg:2013pbi}, depending on the driving amplitude and instability thresholds. A quantitative assessment of locking therefore requires the overlap and damping rate of the same mode in a specified stellar model.  In general though, poorer internal dissipation should lead to larger torques when the mode is fully saturated.

Let us now compare the tidal torque to the gravitational-wave induced torque.  For gravitational waves, the relevant torque is 
\begin{align}
    \frac{|\tau_{\GW}|}{|E_{\orb}|} &\sim \frac{1}{\Omega_{\orb}}\frac{\dot E_{\GW}}{E_{\orb}} = \frac{1}{\Omega_{\orb}t^{(\GW)}_{\rm inspiral}} \\ & \sim 5\e{-7} \frac{\eta}
    {10^{-3}} \quant{\frac{x}{0.0175}}^{5/2},
\end{align}
with $x$ again the post-Newtonian parameter defined in Eq.~\eqref{eq:x-PN}. A necessary condition for resonance locking is that the maximum available tidal torque be at least as large as the torque required by \mainEqTorque, with the appropriate sign. Whether this condition is satisfied depends on the mode overlap and damping rate, together with the orbital and stellar moments of inertia. For low-order $g$-modes, e.g. at $\sim 50\,\Hz$, the mode overlap may be large enough that the necessary condition is possibly satisfied directly. For high-order g-modes, $\sim 1\,\Hz$, nonlinearities would likely be necessary to generate sufficiently large tidal overlaps.

It's also possible that additional torques act on the system which modify the dynamics.  For example, electromagnetic spin-down torques may be important in normal pulsar systems.  In this case, the torque balance as written in the main text would not be correct, although resonance locking may still be possible.  In this case, though, magnetic field drag on the black hole should also be incorporated, which is nontrivial to compute.  We leave this to future work.

The tidal torque expression Eq.~\eqref{eq:estimate-tau-tide-dyn} assumes that saturation of the resonant instability is possible in the time before gravitational radiation drives the system out of resonance.  This assumption hinges on a slow inspiral, and thus on a \emph{small} symmetric mass ratio $\eta$ (equivalently, a small companion mass relative to the neutron star). Therefore, despite the fact that both torques scale with the symmetric mass ratio $\eta$, the tidal torque implicitly scales much more strongly with $\eta$ because mode saturation will be possible only when $\dot\Omega_{\rm orb}/\Omega_{\rm orb}^2\ll\gamma_\alpha/\Omega_{\rm orb}$.  Strictly, to fully saturate the resonance within the pure linear model, $\dot \Omega_{\orb} \sim\gamma_{\alpha}^2$.  However, we assume the resonance saturates well before reaching the maximum amplitude through another, likely nonlinear, mechanism.  It's also possible that other processes amplify the mode amplitude transiently (such as a starquake), or slow down the inspiral transiently (such as a mean-motion resonance with another satellite), to allow locking.  Nonetheless, the derivations we use here will not easily apply to near-equal mass binaries, unless the mode frequency is surprisingly fine-tuned to the NS spin frequency, so $\Omega_{\orb}$ at locking is $\ll 1\,\mathrm{Hz}$. With a dense NS g-mode spectrum, though, this could nonetheless happen for some small number of equal mass binaries.  If mode quality factors are lower, then saturation happens much faster, though the maximum torque the BH can provide in the locked phase is smaller as well.

Saturation of the resonant tidal energy transfer will occur before $I_{\NS} =  I_{\rm orb}/3$.  There exist some general benchmarks for when the resonance-lock condition must break down.    Nonlinearities very likely become important when the modes are driven to such large amplitudes; we give a discussion of one possible nonlinear effect in the next section.

\subsection{Dynamics of the system near resonance}
When the resonance condition $\omega_X = m(\Omega_{\NS}-\Omega_{\orb})$ is satisfied, the companion's tidal field resonantly excites the neutron star's $X$-mode.
In the absence of rotation, a mode of an isolated star with frequency $\omega_X$ is associated with a solution to the perturbed Einstein equations of the form 
\begin{equation}
    \vec \xi(t, r, \theta, \phi)\propto \exp\quantb{i\quant{\omega_Xt -m\phi}} \vec \xi(r, \theta)\,,
\end{equation}
where typically $\vec{\xi}$ is the Lagrangian displacement vector of the fluid. Using a model for resonant tidal excitation from~\citet{Ho:1998hq}, and in the absence of external forces, the Newtonian equations of motion for the mode amplitude are 
\begin{multline}   
    \label{eq:driven-oscillator}
    \ddot a_{X}(t) - 2 i T_{X} \dot a_X +(\omega^2_X- 2T_X \omega_X)a_X  =\\ \frac{M_{\BH}}{r^3}\quantb{\hat Q_{\ell m, \alpha}e^{i(-m\phi-\omega_X t)} -\hat Q_{\ell-m, \alpha}e^{i(m\phi-\omega_X t) }},
\end{multline}
where $a_X$ is the complex mode amplitude and $\hat Q_{\ell m}$ represents\footnote{This is $W_{\ell m} Q_{\ell m, \alpha}$ in the notation of~\cite{Ho:1998hq}.  We assume the neutron star spin is aligned with the orbit, which allows for several simplifications.} the overlap of the mode with the tidal field. Finally $T_X$ represents the rotational contributions to the dynamics,
\begin{equation}
    T_X=  -m\Omega_{\NS}  + i \int d^3x \rho \vec\xi_X\cdot\quant{\vec \Omega_{\NS} \times \vec \xi_X}.
\end{equation}
In this work, we follow \cite{Fuller:2016} and assume the inertial contributions to the equations of motion are negligible, $T_X=0$.  This is, in general, not a good assumption, but we expect our overall picture of resonance locking to remain unchanged by the introduction of inertial effects (though inertial modes may be an additional source of resonances, which may or may not lock efficiently~\cite{Fuller:2016}). 

We do wish to model weak interactions with other system degrees of freedom, which can be taken to include inertial effects, damping, and  nonlinearities (potentially including nonlinear interactions with other modes). We must therefore add to the right-hand side of Eq.~\eqref{eq:driven-oscillator} an external force of the form
\begin{equation}
    f_{\rm ext} = f_{\rm ext}(t, a_X, \dot a_X, \dots).
\end{equation}
Schematically then, Eq.~\eqref{eq:driven-oscillator} becomes
\begin{equation}
    \ddot a_X  + \omega_X^2a_X = A_X e^{im\Phi(t)} + f_{\rm ext},
\end{equation}
where $A_X(r) = M_{\rm BH}/r^3\hat Q_{\ell m, X}$, $\Phi'(t) = \phi'(t) - \Omega_{\NS}(t)$, and we have assumed the system is close to resonance, so $\omega_X = m\dot \Phi$ and the second term from Eq.~\eqref{eq:driven-oscillator} is dominant. This simplified model captures the key details of the problem we are interested in.

We can analyze this system in the slowly varying envelope approximation, setting 
\begin{equation}
    a_X = c_Xe^{i\omega_Xt}
\end{equation}
to find
\begin{equation}
    2i\omega_X\dot c_X = 
    A_X e^{im\Phi(t)-i\omega_X t} + f_{\rm ext}e^{-i \omega_Xt}.
\end{equation}
For simplicity, we now consider a model for $f_{\rm ext}$ which will allow the problem to be analyzed 
\begin{equation}
    f_{\rm ext} = -2\gamma \dot a_X - \beta a_X^3,
\end{equation}
where $\gamma$ and $\beta$ should be small enough that these terms are unimportant when the amplitude of the system is small.  
The envelope equation then becomes
\begin{equation}
    2i\omega_X\dot c_X = A_X e^{i\delta \phi(t)}  - 2i\gamma \omega_X c_X -  \frac{3}{4}\beta |c_X|^2 c_X.
\end{equation}
Further decomposing $c_X = Re^{i\theta}$, one finds
\begin{align}
    \dot R &= -\gamma R + \frac{A_X}{2\omega_X}\sin(\theta-\delta \phi);\\
    \dot \theta &= \frac{3\beta}{8\omega_X} R^2 + \frac{A_X}{2\omega_XR}\cos(\theta-\delta \phi),
\end{align}
so the nonlinear term modifies the resonance condition via a frequency shift 
\begin{equation}
    \omega_X' = \omega_X + \frac{3\beta}{8\omega_X}R^2.
\end{equation}

The essential point of this analysis is that even if the resonance lock is maintained, nonlinear terms will eventually perturb the resonance condition slightly.  In general, perturbations that push the system further into lock will lead to larger mode amplitudes and stronger nonlinear corrections.  In this simple  model, the mode resonance is pushed to higher frequency, so it is achieved at smaller orbital radius.  There is a natural feedback mechanism in this system, where rapid dissipation from nonlinear terms will spin up the neutron star, while, on the other hand, slow dissipation will lead to larger mode amplitudes and stronger nonlinear corrections.  The only effect that may change this qualitative picture is the emergence of a parametric instability. Still, we expect the emergence of such an instability to saturate and quickly lead to a large amount of internal dissipation in the star, which would also result in spin modifications.  In general, nonlinear effects in the star will tend to lead to more possible relevant degrees of freedom which can couple to the tidal force, but also more possible mechanisms to dissipate stored mode energy.

\end{document}